\documentclass[a4paper]{jpconf}
\usepackage{graphicx}
\begin{document}
\title{The CDMS view on molecular data needs of \textit{Herschel}, SOFIA, and ALMA}

\author{Holger S~P M\"uller, C~P Endres, J Stutzki and S Schlemmer}

\address{I.~Physikalisches Institut, Universit\"at zu K\"oln, Z\"ulpicher Str.~77, 
         50937~K\"oln, Germany}

\ead{hspm@ph1.uni-koeln.de}

\begin{abstract}
The catalog section of the Cologne Database for Molecular Spectroscopy, CDMS, 
contains mostly rotational transition frequencies, with auxiliary information, 
of molecules observable in space. The frequency lists are generated mostly from 
critically evaluated laboratory data employing established Hamiltonian models. 
The CDMS has been online publicly for more than 12~years, e.g., via the short-cut 
\verb"http://www.cdms.de". Initially constructed as ascii tables, its inclusion 
into a database environment within the Virtual Atomic and Molecular Data Centre 
(VAMDC,  \verb"http://www.vamdc.eu") has begun in June 2008. A test version of 
the new CDMS is about to be released. The CDMS activities have been part of the 
extensive laboratory spectroscopic investigations in Cologne. Moreover, these 
activities have also benefit from collaborations with other laboratory spectroscopy 
groups as well as with astronomers. We will provide some basic information on 
the CDMS and its participation in the VAMDC project. In addition, some recent 
detections of molecules as well as spectroscopic studies will be discussed 
to evaluate the spectroscopic data needs of \textit{Herschel}, SOFIA, and ALMA 
in particular in terms of light hydrides, complex molecules, and metal containing 
species. 
\end{abstract}

%%%%%%%%%%%%%%%%%%%%%%%%%%%%%%%%%%%%%%%%%%%%%%%%%%%%%%%%%%%%%%%%%%%%%%%%%%%%%%%%%%%%%%%%%
%%%%%%%%%%%%%%%%%%%%%%%%%%%%%%%%%%%%%%%%%%%%%%%%%%%%%%%%%%%%%%%%%%%%%%%%%%%%%%%%%%%%%%%%%
%%%%%%%%%%%%%%%%%%%%%%%%%%%%%%%%%%%%%%%%%%%%%%%%%%%%%%%%%%%%%%%%%%%%%%%%%%%%%%%%%%%%%%%%%

\section{Introduction}
\label{Introduction}

The assignment of emission or absorption features observed by radio astronomical means to 
an atomic or molecular species relies on the availability of rest frequencies with sufficient 
accuracy. Usually, and in particular in the case of identifying a newly detected molecule, 
such data are taken or derived from laboratory spectroscopic data, as in the case of the 
CF$^+$ cation, a molecule which is isoelectronic to CO and which was detected for the first time 
in the photon-dominated region (PDR) Orion Bar \cite{det-CF+_2006}. We should mention, however, 
that rest frequencies may also be taken from radio astronomical observations. In rare cases, 
a new molecular species may even be identified from such data in the absence of laboratory 
data, but in combination with quantum chemical calculations, as in the case of C$_3$H$^+$, 
which has been detected very recently in the Horsehead PDR in Orion \cite{det-C3H+_2012}. 
It should be noted that some scientists view the detection as tentative because of the 
absence of confirming laboratory spectroscopic data.

The Cologne Database for Molecular Spectroscopy, CDMS, has been created more than 12 years 
ago to provide in its catalog section rest frequencies of molecules as well as some atoms 
which have been or may be detected in space by means of radio astronomy. The CDMS has been 
available online free of charge at \verb"http://www.astro.uni-koeln.de/cdms/"; 
the short-cut \verb"http://www.cdms.de" may also be used. Background information may be 
obtained in each section of the CDMS. Additional information as well as up-to-date accounts 
have been published some time ago \cite{CDMS_1,CDMS_2}. One or both should be cited 
to acknowledge the use of the CDMS. Some important aspects of the CDMS as well as 
the transformation of its catalog section from \verb"ascii" tables into a Virtual Observatory 
compliant database environment within the framework of the Virtual Atomic and Molecular 
Data Centre (VAMDC) will be dealt with in section~\ref{cdms-section}.

The CDMS activities have been part of the much longer ongoing laboratory spectroscopic 
investigations in Cologne which involve mostly rotational spectroscopy in the millimeter 
to terahertz regions in support of observational astronomy. The CDMS not only acts as 
a link between laboratory spectroscopy and radio astronomy, but it also benefits from 
collaborations with corresponding groups. We will present data needs from the CDMS' 
point of view of terahertz missions, such as the \textit{Herschel Space Observatory}, 
in section~\ref{herschel} and of interferometer arrays, such as the Atacama Large Millimeter 
Array (ALMA), in section~\ref{alma}. As the data needs have been evaluated considering 
observational as well as laboratory spectroscopic results, recent findings in either area 
will be summarized.

%%%%%%%%%%%%%%%%%%%%%%%%%%%%%%%%%%%%%%%%%%%%%%%%%%%%%%%%%%%%%%%%%%%%%%%%%%%%%%%%%%%%%%%%%
%%%%%%%%%%%%%%%%%%%%%%%%%%%%%%%%%%%%%%%%%%%%%%%%%%%%%%%%%%%%%%%%%%%%%%%%%%%%%%%%%%%%%%%%%
%%%%%%%%%%%%%%%%%%%%%%%%%%%%%%%%%%%%%%%%%%%%%%%%%%%%%%%%%%%%%%%%%%%%%%%%%%%%%%%%%%%%%%%%%

\section{The Cologne Database for Molecular Spectroscopy}
\label{cdms-section}

%%%%%%%%%%%%%%%%%%%%%%%%%%%%%%%%%%%%%%%%%%%%%%%%%%%%%%%%%%%%%%%%%%%%%%%%%%%%%%%%%%%%%%%%%

\subsection{The CDMS catalog}
\label{catalog}

The catalog section (\verb"http://www.astro.uni-koeln.de/cdms/catalog") of the CDMS is 
the most important one for radio astronomers. It provides line lists of transition 
frequencies for molecular species which have been or may be observed in space. 
The frequency ranges covered range mostly from radio frequencies (below $\sim$3~GHz or 
wavelengths greater the 10~cm) over microwave and millimeter wave to the submillimeter wave 
and terahertz (or far-infrared) regions, as far as this may be useful, and concern for 
the most part rotational transitions with fine or hyperfine structure splitting as far as 
appropriate. Atomic fine structure transitions as well as rotation-vibration transitions 
are available to some extent, and the frequencies may extend into the mid- or even 
near-infrared regions in these cases. Information on the catalog section is available at 
the internet address given above as well as in two publications \cite{CDMS_1,CDMS_2}. 

The information on each molecular or atomic species are contained in one individual entry, 
and separate entries have been generated for different isotopic species and usually for 
excited vibrational states, as far as these may be of relevance for radio astronomy. 
As of October 2012, there have been 657 different entries in the CDMS catalog of which 
at least 279 have been detected in the interstellar medium (ISM) or in circumstellar 
envelopes (CSE) of late-type stars. These 279 species represent a very large fraction 
of the more than 170 different molecules which have been identified unambiguously 
in the ISM or in CSE. Around 5 entries are created newly or updated each month with a 
very large scatter mainly because the generation of an entry may take from about one hour 
to considerably more than one month. Not only is the content of the CDMS catalog growing 
fairly rapidly, but also the number of users increases also. Currently, there are around 
2000 to 2500 accesses to the CDMS catalog each month which correspond to around 500 
different users. 

The CDMS follows the tradition set by the JPL millimeter and submillimeter catalog
\cite{JPL-catalog_1998}. In fact, one of the authors of this work (HSPM) 
has contributed to the JPL catalog, and one of the incentives to found the CDMS was 
due to the limitations to contribute to the JPL catalog from outside. 

Each of the more than 650 entries is currently an \verb"ascii" table with anywhere between 
one line and several ten thousands of lines. Each of the 80 character lines represents 
one spectral feature which may be observed in emission or in absorption. It contains 
essentially all of the information necessary to analyze radio astronomical observations, 
namely the frequency and its uncertainty in megahertz units, the absorption intensity 
at 300~K, the degree of freedom in the rotational partition function, the lower state 
energy in inverse centimeters, the upper state degeneracy, the tag, a code for the format 
of the quantum numbers, and finally the upper state and the lower state quantum numbers. 
The absorption intensity at 300~K is generally not particularly useful for the intended 
purposes of the CDMS, however, this is due to the program which is most commonly used 
to generate entries and which was initially written to support investigations of 
Earth's atmosphere. A search and conversion routine is available to convert the intensity 
at 300~K to that at selected lower temperatures, to the Einstein $A$ value, or to the 
line strength $S \mu^2$. The tag is used to designate an entry. It consists of the 
molecular weight of the species followed by a 5 to differentiate the CDMS catalog entries 
from those in the JPL catalog, in which an equivalent scheme is employed, followed by 
two counting digits for species with the same molecular weight. The limitation of 80 
characters restricts the number of quantum numbers to 6 for both upper and lower state 
and to two characters for each quantum number. In addition, half-integer quanta are rounded 
up, and quantum numbers larger than 99 are indicated by using the alphabet for the tens 
position (A stands for 10, B for 11, and so on). 

The entries are generally created by fitting laboratory data to established Hamiltonian 
models. Rest frequencies from astronomical observations may be used also. One of the most 
important requirements for the resulting entries to be as reliable as possible is the 
critical evaluation of the rest frequencies used in the modeling. It is not so uncommon 
that experimental uncertainties are not reported or they are too optimistic or too 
pessimistic. And even if uncertainties appear to be adequate overall there may be still 
issues, such as misassignments or typographical errors etc, in particular in large 
or divers data sets. It is also mandatory to evaluate Hamiltonian models. If rest 
frequencies for a certain entry come from just one publication and if their number as 
well as that of the spectroscopic parameters is small, this is usually straightforward, 
and problems with a Hamiltonian model are rather rare. The task may be much more complex 
if the set of rest frequencies is large or if they come from multiple publications, 
even more so as it is fairly common that none of the publications deal with a sufficiently 
large subset of the rest frequencies. Moreover, we have found fairly frequently that 
Hamiltonian models may be incomplete or too extensive or even not appropriate at all. 
An interesting case in this respect is a recent publication of rest frequencies 
determined for the asymmetric rotor molecule NaCN. The Hamiltonian was diagonalized 
incompletely such that a very large set of spectroscopic parameters did not achieve 
reproduction of the transition frequencies within experimental uncertainties. 
The complete diagonalization of the Hamiltonian reproduced the transition frequencies 
well within experimental uncertainties with a set of spectroscopic parameters much 
smaller \cite{NaCN-spec_2012} than initially \cite{NaCN-spec_2011}. It is quite clear 
that experienced spectroscopists are needed to generate entries as reliable as possible.  

The entries have been generated for the most part with the \verb"spfit"/\verb"spcat" 
program suite which is rather versatile, permitting linear, symmetric, as well as 
asymmetric rotors to be treated and may take vibration-rotation interaction or 
electron spin or nuclear spin-interaction into account \cite{JPL-catalog_1998,spfit_1991} 
and which has evolved over the years \cite{ed_Cohen-Pickett_2008,JPL-catalog_2010}. 
Quantum number assignments in the CDMS catalog are based entirely on Hunds's case (b), 
and this applies to the JPL catalog for the most part. The quantum numbers are (as far as needed) 
the total rotational quantum number $N$ with its projection $K$ and total parity for symmetric 
top quantum number labeling or $N$, $K_a$, and $K_c$ for asymmetric top quantum number labeling; 
followed by (vibrational) state identifiers and spin-quanta, with $F$ given last. Additional 
information is available on the first page of the CDMS catalog and in the documentation file 
\verb"spinv.pdf" generated by H.~M. Pickett. Other quantum number assignments may be generated 
as described in the next subsection. 

There are several auxiliary files available. The most important one is the documentation 
file which provides some basic background information on the species and on the entry. 
This includes information on who created the entry when, the version number, source(s) 
of the transition frequencies used to generate the entry with potential comments to 
the frequencies or to the Hamiltonian employed. Also given are partition function values 
at specific temperatures, possibly comments on the calculation of the partition function 
and estimates on the reliability of the entry. The documentation may also contain links 
to special calculations, e.g. with hyperfine structure if this is not needed throughout, 
or to calculations for selected spin-states, such as {\it ortho} or {\it para}. 
Energy files which contain at least all of the levels in the entry should also be available 
for all entries. The line, parameter, and intensity files, which have been used to generate 
the entry, are frequently available in the archive section of the catalog or 
in other sections of the CDMS.

%%%%%%%%%%%%%%%%%%%%%%%%%%%%%%%%%%%%%%%%%%%%%%%%%%%%%%%%%%%%%%%%%%%%%%%%%%%%%%%%%%%%%%%%%

\subsection{The CDMS within the VAMDC framework}
\label{vamdc}

The CDMS catalog in its current form is a resource considered to be useful by many astronomers, 
in particular those dealing with \textit{Herschel} observations. However, the organization of 
the entries in \verb"ascii" tables has not only increasing disadvantages the more the CDMS 
catalog grows, but it is also not well suited for the exchange of data. Therefore, we have 
started to put the CDMS catalog into a database environment. This has started in the second 
half of 2009 in a concerted effort within the Virtual Atomic and Molecular Data Centre 
(VAMDC, \verb"http://www.vamdc.eu") \cite{VAMDC_2010}, which is a project within the 
European Union Framework Programme 7. 15 partners from the EU and Russia plus two 
external partners from the USA cooperate to provide common infrastructure to access 
the $\sim$20 different spectroscopic, collisional, and kinetic databases. 

Common standards (\verb"http://www.vamdc.eu/standards"), such as data exchange formats, 
query languages, and protocols have been defined and applied to a large number of databases, 
including the CDMS. This allows to query all databases in a uniform way by simple 
url-requests and to retrieve the data in an XML format called VAMDC-XSAMS. A web-portal 
has been released (\verb"http://portal.vamdc.eu") where the data of these databases 
can be accessed. Care has to be taken as some of the participating databases, including 
the CDMS, have not been released these services officially because quality control 
of the data has not been finished yet. In addition, efforts are under way to include 
further partners from additional countries worldwide.

The database version of the CDMS catalog will not only provide the current access options, 
browse selected entries, but there will be further options in the search and conversion 
routine, e.g. other frequency or intensity units or quantum number assignments following 
Hund's case (a). This includes also a new unit ''molecule'' which comprises all isotopologs 
and all vibrational states pertaining to this molecule. The CDMS will be accessible from 
all participating databases, and requests may be sent through the CDMS portal or via URL. 
Various types of output formats will be available, and particular care has been taken 
to use well-defined and well explained quantum numbers, irrespective of the difficulties 
to uniquely assign quantum numbers in not so rare cases such as interactions of various 
kind.

%%%%%%%%%%%%%%%%%%%%%%%%%%%%%%%%%%%%%%%%%%%%%%%%%%%%%%%%%%%%%%%%%%%%%%%%%%%%%%%%%%%%%%%%%
%%%%%%%%%%%%%%%%%%%%%%%%%%%%%%%%%%%%%%%%%%%%%%%%%%%%%%%%%%%%%%%%%%%%%%%%%%%%%%%%%%%%%%%%%
%%%%%%%%%%%%%%%%%%%%%%%%%%%%%%%%%%%%%%%%%%%%%%%%%%%%%%%%%%%%%%%%%%%%%%%%%%%%%%%%%%%%%%%%%

\section{Data needs of terahertz missions}
\label{herschel}

\subsection{Missions}

Radio astronomical observations at terahertz frequencies are challenging because Earth 
atmospheric water vapor in particular and oxygen as well as ozone to a lesser extent 
block large portions of these frequencies or reduce the transmission considerably. 
One way to overcome these challenges is to deploy an observatory outside of Earth's 
atmosphere. The \textit{Herschel Space Observatory} is a mission of the European Space 
Agency (ESA) with science instruments provided by European-led Principal Investigator 
consortia and with important participation from NASA. Its initial name was Far Infrared 
and Sub-millimetre Telescope (FIRST), but is was renamed in 2000 to honor the eminent 
astronomers Wilhelm/William and Caroline Herschel. It was launched 14 May 2009, and it 
should run out of liquid helium as coolant around March 2013. There are three different 
instruments on board of the \textit{Herschel} satellite which is equipped with a 3.5~m 
telescope. The most important one from the spectroscopic point of view is the Heterodyne 
Instrument for the Far-Infrared (HIFI). It is a high-resolution ($\sim$10$^6$ or $\sim$10$^7$) 
instrument which covers 480$-$1250~GHz and 1410$-$1910~GHz. The Spectral and Photometric 
Imaging REceiver (SPIRE) and the Photodetector Array Camera and Spectrometer (PACS) can act 
as photometers as well as moderate resolution spectrometers. They cover 194$-$671~$\mu$m and 
57$-$210~$\mu$m, respectively, corresponding to 447$-$1550~GHz and $\sim$1430$-$5260~GHz, 
respectively. 152 letters in volume 518 of \textit{Astronomy and Astrophysics} deal with 
early \textit{Herschel} results; volume 521 presents 50 plus 2 letters of results 
obtained with HIFI. 

The Stratospheric Observatory For Infrared Astronomy (SOFIA) is a mission lead by NASA 
with German participation through the Deutsches Zentrum f\"ur Luft- und Raumfahrt (DLR), 
which is the German space organisation. It has a 2.7~m telescope on board of a Boing~747SP 
airplane which is capable of reaching altitudes of 18~km. The most important advantage 
with respect to \textit{Herschel} is the possibility of not only making modifications 
to the existing instruments, but also to develop new instruments. The disadvantages of 
such an airplane mission are quite clear also: it has a lot of O$_2$, most of O$_3$, 
and, more importantly, still a fair amount of H$_2$O in the optical path, which affects 
or even blocks certain frequencies. Moreover, a single source can only be observed for 
a fair fraction of a flight which will last 11~h at most. SOFIA saw first light on 
April 1, 2011. One of the two early science instruments is a high-resolution 
heterodyne spectrometer: the German REceiver for Astronomy at Terahertz frequencies 
(GREAT). It has currently three wider frequency channels covering 1252$-$1392, 
1417$-$1520, and 1815$-$1910~GHz as well as one narrow channel around the OH ground 
state transition near 2512~GHz. Two additional narrow channels around the HD 
$J = 1 - 0$ transition near 2675~GHz and around the fine structure ground state 
transition of atomic oxygen around 4745~GHz are under construction, further channels 
are being developped. Early results obtained with GREAT have been published in 22 
letters in volume 542 of \textit{Astronomy and Astrophysics}.

Terahertz astronomy is also possible from ground, albeit with even lower transparency 
of the atmosphere than available with SOFIA. The Atacama Pathfinder EXperiment (APEX) 
with its 12~m telescope, which is a modified ALMA (see section~\ref{alma}) prototype 
located in the vicinity of ALMA, is particularly worth mentioning. The project is a 
collaboration between the Max-Planck-Institut f\"ur Radioastronomy (MPIfR) in Bonn, 
Germany, with the Onsala Space Observatory (OSO) and the European Southern Observatory 
(ESO). It covers several atmospheric windows between about 200 and 1600~GHz. 
Early results obtained with APEX have been published in 102 letters in volume 454 of 
\textit{Astronomy and Astrophysics}. There have been radio telescopes with submillimeter 
capabilities before, but these have been more limited in terms of sensitivity or 
highest frequency achievable at least somewhat routinely. The Caltech Submillimeter 
Observatory on Mauna Kea, Hawaii, with its 10.4~m dish is worthwhile mentioning, 
but its decommissioning is planned to start in 2016.

%%%%%%%%%%%%%%%%%%%%%%%%%%%%%%%%%%%%%%%%%%%%%%%%%%%%%%%%%%%%%%%%%%%%%%%%%%%%%%%%%%%%%%%%%

\subsection{Goals and findings of terahertz missions}
\label{findings-Herschel}

One of the major goals of the \textit{Herschel} mission was actually the study of the 
chemical complexity of various types of sources, such as hot dense molecular clouds, 
cold dense molecular clouds, PDRs, the envelopes of late type stars, and near-by galaxies.
One way of achieving this goal were unbiased line surveys covering the entire frequency 
range of one or more instruments.

The goals of species to be studied spectroscopically by \textit{Herschel} are probably 
the most general ones; those of GREAT or APEX have to take into account the transmission 
properties of Earth's atmosphere, and the different frequency regions available will 
affect their goals somewhat also.

Among the most importants species to be studied were O$_2$ and low energy transitions 
of H$_2$O because these cannot be studied from the ground and not even by SOFIA. 
Other very important goals include atomic fine structure lines, most notably those 
of C$^+$, O, and N$^+$, (especially fundamental) rotational transitions of light 
hydrides of the type AH$_{\rm n}$ or AH$_{\rm n} ^+$, and higher energy lines of 
abundant molecules such as CO, HCN, HNC, HCO$^+$. Other important goals include the 
low-lying bending modes of C$_3$ and possibly other carbon-chain molecules without 
permanent dipole moment as well as complex molecules.

Emission features of O$_2$ in the interstellar medium have been found with \textit{Herschel} 
beyond any doubt for the first time, however, they were very weak, indicating very low 
column densities \cite{O2_2011,O2_2012}. H$_2$O and its isotopologs have been studied 
extensively with \textit{Herschel}, and the abundances were found to vary very greatly. 
Other previously detected light hydrides have been observed also very frequently; these 
include CH$^+$, CH, NH$_3$, NH$_2$, NH, OH, H$_3$O$^+$, and HF.
Complex molecules turned out to be more difficult to observe than some astromers had hoped. 
Nevertheless, features of methanol and dimethyl ether have been detected in all or 
almost all of the HIFI channels and features of ethyl cyanide, vinyl cyanide, and methyl 
formate were easily recognizable in the lower frequency channels (up to 1250~GHz). 
Rather unexpected were the extensive and in part strong emission lines of SiC$_2$ 
seen even well beyond 1~THz in the circumstellar envelope of the late type star CW~Leonis, 
also known as IRC~+10216. As the amount of published laboratory data was comparatively 
sparse and extended only to 370~GHz, it was possible to improve the spectroscopic 
parameters of this molecule considerably by including the astronomical data in a 
combined fit \cite{SiC2_2012}. 

Probably the most remarkable spectroscopic findings of recent terahertz observations 
are the first observations of six light hydrides and of two deuterated isotopologs. 
The cations OH$^+$ \cite{det-OH+_2010} and SH$^+$ \cite{det-SH+_2011} have been detected 
in absorption toward the prominent Galactic center source Sagittarius~B2(M) with APEX 
from the ground. Both molecules, as many other light hydrides, showed also absorption 
features toward intervening diffuse spiral arm clouds in their fundamental transitions. 
Subsequently, both were also seen with the \textit{Herschel Space Observatory}, OH$^+$ 
even in several Extragalactic sources. H$_2$O$^+$ \cite{det-H2O+_2010} and H$_2$Cl$^+$ 
\cite{det-H2Cl+_2010} were detected with \textit{Herschel}-HIFI, and they were also seen 
toward many other sources in addition to the ones in the detection letters. The HCl$^+$ 
radical was detected with HIFI in absorption toward the Galactic star-formation regions 
W31C and W49N \cite{det-HCl+_2012}, and the isoelectronic SH radical was detected with 
GREAT on board of SOFIA toward the latter source \cite{det-SH_2012}. The deuterated 
radicals ND \cite{det-ND_2010} and OD \cite{det-OD_2012} were detected in absorption 
toward the solar-mass protostar IRAS~16293-2422 with HIFI and GREAT, respectively. 

Several light hydride species have been detected with \textit{Herschel} in external 
galaxies, among them OH$^+$ \cite{Mrk_231_OH+_2010} and H$_2$O$^+$ \cite{M82_H2O+_2010}. 
Excitation conditions in external galaxies may be quite different from those commonly 
found in Galactic sources. Highly excited OH$^+$ and H$_2$O$^+$, as well as H$_3$O$^+$, 
in NGC~4418 and Arp~220 were reported in the course of the reviewing process of this article 
\cite{H1-3O+_Arp220_2013}. Such observations extend the need for laboratory spectroscopic 
investigations to higher quantum numbers.

%%%%%%%%%%%%%%%%%%%%%%%%%%%%%%%%%%%%%%%%%%%%%%%%%%%%%%%%%%%%%%%%%%%%%%%%%%%%%%%%%%%%%%%%%

\subsection{Recent laboratory spectroscopic investigations into light hydrides}

Federman et al \cite{IAU-report_2011} provide information on laboratory spectroscopic studies 
which may be important for astronomers and which have been published between the second half 
of 2002 and the first half of 2011. Here we summarize recent (2008 and after) studies on 
light hydrides which are necessary or may be useful for studies of these molecules with 
terahertz missions such as \textit{Herschel}. 

The spectra of ionic light hydrides are often particularly difficult to obtain, and 
they are usually very sparse and sometimes very complex. Therefore, it is not surprising 
that on occasion incorrect transition frequencies enter the scientific literature. Two 
noteworthy examples concern the $J = 1 - 0$ transitions of H$_2$D$^+$ \cite{H2D+_rot_2008} 
and CH$^+$ \cite{CH+_rot_2010} which were measured recently more than 60~MHz away from 
previous measurements in both cases. In addition, the $J = 1 - 0$ transitions of $^{13}$CH$^+$ 
and CD$^+$ were measured for the first time \cite{CH+_rot_2010}. In conjunction with data 
from electronic spectra, extrapolation to higher quantum numbers became possible 
\cite{CH+_analysis_2010}. Additional information on the energy levels of H$_2$D$^+$ is 
available from measurements of several other rotational transitions with greatly differing 
accuracies as well as from measurements in the infrared region.

Absorption features of HCl$^+$ \cite{det-HCl+_2012} could only be accounted for properly 
after new and much more accurate laboratory measurements had been carried out 
\cite{HCl+_rot_2012}. The first rotational transitions, mostly of low energy, were obtained 
for CH$_2$D$^+$ \cite{CH2D+_rot_2010}, for H$_2$F$^+$ \cite{H2F+_rot_2011,H2F+_rot_2012}, 
and for the isotopolog $^{15}$NH \cite{15NH_rot_2012}. However, the prospects of observing 
CH$_2$D$^+$ are uncertain because of the very low dipole moment and for H$_2$F$^+$ because 
of the possibly unfavorable chemistry. Transitions mainly involving higher energy levels 
were recorded for NH$_3$ \cite{NH3_rot_2010,HD_H2O_NH3_2011} and for H$_2$O 
\cite{HD_H2O_NH3_2011,H2O_rot_2012}. Additional or more accurate data or improved analyses 
were reported for CH$_3$D \cite{CH3D_2009}, H$_3$O$^+$ \cite{H3O+_rot_2009}, H$_2$DO$^+$ 
\cite{H2DO+_analysis_2010}, SH$^+$ \cite{SH+_analysis_2009}, SH \cite{SH_rot_2011}, and 
HD \cite{HD_H2O_NH3_2011}.

%%%%%%%%%%%%%%%%%%%%%%%%%%%%%%%%%%%%%%%%%%%%%%%%%%%%%%%%%%%%%%%%%%%%%%%%%%%%%%%%%%%%%%%%%

%%%%%%%%%%%%%%%%%%%%%%%%%%%%%%%%%%%%%%%%%%%%%%%%%%%%%%%%%%%%%%%%%%%%%%%%%%%%%%%%%%%%%%%%%

\begin{figure}
\begin{center}
\includegraphics[angle=0,width=15cm]{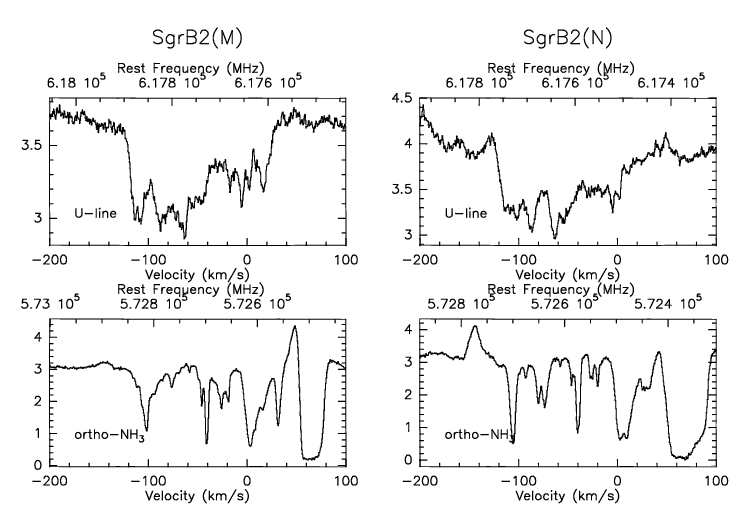}
\end{center}
\caption{\label{dib}Absorption features observed with the HIFI instrument on board of the 
         \textit{Herschel} satellite toward the giant Galactic molecular clouds Sagittarius~B2(M) 
         (left) and Sagittarius~B2(N) (right). The upper traces show absorption of an unidentified 
          species, the lower traces show absorption of the ground state $J = 1 - 0$ transition 
          of NH$_3$. Note: whereas NH$_3$ shows largely saturated absorption at velocities of 
          around 70~km/s, which are due to the sources Sgr~B2(M) and Sgr~B2(N), respectively, 
          there is no or essentially no such absorption visible for the unidentified species. 
          It only shows absorption in the diffuse spiral arm clouds between the sources 
          and Earth which occur at velocities between about $-$120 and 40~km/s, which is also 
          seen in the case of NH$_3$. A small part of the absorption is caused by a transition 
          of methyleneimine. The figure was provided by P. Schilke. It is based on data from two 
          sub-programs within the HEXOS key program \cite{HEXOS_2010} which are lead by P. Schilke 
          and D.~C. Lis for Sgr~B2(M) and Sgr~B2(N), respectively. The data are unpublished thus far.}
\end{figure}

%%%%%%%%%%%%%%%%%%%%%%%%%%%%%%%%%%%%%%%%%%%%%%%%%%%%%%%%%%%%%%%%%%%%%%%%%%%%%%%%%%%%%%%%%

\subsection{Data needs}

Light hydride species remain important subjects in terms of data needs of terahertz 
astronomy even though a large number of them have been detected in space recently 
and new and improved rest frequencies have become available for several because of 
laboratory measurements. Accurate data for the lowest energy transitions of CH$_2 ^+$ 
and possibly H$_2$S$^+$ appear to be most urgently needed even if the prospects of 
detecting these molecules differ greatly in model calculations. The explanation 
is the following: in the reaction chain X$^+$ + H$_2$ $\rightarrow$ HX$^+$ + H, 
HX$^+$ + H$_2$ $\rightarrow$ H$_2$X$^+$ + H, H$_2$X$^+$ + H$_2$ $\rightarrow$ H$_3$X$^+$ + H, 
the first reaction is highly endothermic at low temperatures for X being C or S, 
and in contrast to, e.g., X being O. Nevertheless, both CH$^+$ as well as SH$^+$ 
are abundant species in diffuse and translucent molecular clouds. The formation 
of H$_2$S$^+$, however, is more uncertain because the reaction between SH$^+$ and 
H$_2$ is also highly endothermic at low temperatures. There are theories which are 
able to explain the high abundances of CH$^+$ and SH$^+$, but thus far, none has 
proven to be entirely satisfactory. Therefore, the observation or the derivation of 
meaningful upper limits to the column densities of CH$_2 ^+$ should shed some 
light on these important and interesting aspects of astrochemistry. The implication 
of a detection of H$_2$S$^+$ in space is less clear at the moment. H$_2$S$^+$ may 
be observed with \textit{Herschel}/HIFI, but possibly also from ground. The search 
for CH$_2 ^+$ will be conducted best with SOFIA/GREAT. In this context, it is 
worthwhile mentioning that CH$_3 ^+$ cannot be observed by means of radio astronomy 
because as a planar symmetric top molecule it does not possess a permanent dipole 
moment. The rotational spectrum of H$_3$S$^+$ has been characterized sufficiently 
well; it may be searched for best in the upper millimeter region.

The NH$^+$ radical cation has, to our knowledge, not been detected thus far despite 
extensive searches with \textit{Herschel}/HIFI. This may be explained by the comparatively 
small column densities of N$^+$. This, in turn, suggests that NH$_2 ^+$ has small 
column densities also, diminishing the need of laboratory data for this molecule. 
Moreover, because of the very low barrier to linearity of NH$_2 ^+$, the low-energy 
levels of this ion are currently only poorly known, and the corresponding transitions 
are probably very weak.

Potential data needs in terms of metal hydrides will be discussed in the next section 
in the context of late-type stars. Data for further hydride species are not needed from 
the current point of view, but in the case of already detected hydrides additional data 
may be useful in some cases, in particular for ALMA.

Data of complex molecules needed for \textit{Herschel} have been obtained, though 
some may have to be published. The situation is very different for ALMA or other 
interferometer facilities and will be discussed in section~\ref{alma}. 

While C$_3$ has been investigated fairly extensively with \textit{Herschel} as well as 
with other missions via its low-lying bending mode, the study of heavier pure carbon chain 
molecules is hampered by the lack of knowledge of accurate data for their bending modes, 
which provide the best way to identify these, as well as other centro-symmetric molecules 
as they do not possess a permanent dipole moment. However, spectroscopic data are 
not only needed for such molecules, but for other carbon-containg molecules with 
4 to 7 heavy atoms as these may be observable at terahertz frequencies also.

There are also data needs which are rather unspecific. An unidentified absorption feature 
has been detected with HIFI toward several source with strong far-infrared emission 
near 617~GHz, see fig.~\ref{dib}. However, this line appears to be present only in the 
diffuse spiral-arm clouds in the line of sight toward the respective FIR sources; 
it is conspicuously absent toward the sources themselves. It is thus, to the best of 
our knowledge, the first submillimeter Diffuse Interstellar Band (DIB). 
The feature may be caused by an atomic or molecular species, possibly in a metastable 
electronic state. If a rotational transition should cause the absorption then 
it contains probably at most two hydrogen atoms, two heavy atoms, and possibly not more 
than three atoms altogether. We do not exclude the possibility that the feature may 
be caused by a vibration-rotation transition of a fairly large molecule.

%%%%%%%%%%%%%%%%%%%%%%%%%%%%%%%%%%%%%%%%%%%%%%%%%%%%%%%%%%%%%%%%%%%%%%%%%%%%%%%%%%%%%%%%%
%%%%%%%%%%%%%%%%%%%%%%%%%%%%%%%%%%%%%%%%%%%%%%%%%%%%%%%%%%%%%%%%%%%%%%%%%%%%%%%%%%%%%%%%%
%%%%%%%%%%%%%%%%%%%%%%%%%%%%%%%%%%%%%%%%%%%%%%%%%%%%%%%%%%%%%%%%%%%%%%%%%%%%%%%%%%%%%%%%%

\section{Data needs of interferometer arrays}
\label{alma}

\subsection{Interferometry $-$ general considerations}

The Atacama Large Millimeter Array (ALMA) is an international project located in the 
Chilean Andes more than 5000~m above sea level which will consist of 66~telescope dishes. 
54 of them have a diameter of 12~m and may be apart between 150~m and 16~km; 12 closely 
spaced 7~m dishes form the ALMA Compact Array (ACA). 
Although still under construction, several receivers are already available which cover 
many atmospheric windows from 84 to 720~GHz. Eventually, frequencies down to about 
30~GHz and up to almost 950~GHz shall cover all atmospheric windows in between. 
The 112 early science projects of cycle~0 should have been completed by now, and the first 
publications involving ALMA observations have appeared in the scientific literature.

Interferometry provides greatly increased angular resolution. And even though ALMA will be 
the most sensitive array in the millimeter and submillimeter regions with the highest 
angular resolution, other arrays exist at similar and also at lower frequency regions, 
and most of the consideration for ALMA apply also to these interferometers.

The higher angular resolution will be beneficial, e.g., for the study of the clumpy structure
of PDRs. It is not easy to predict which molecules else may be detected in such environments. 
However, recent detections, such as CF$^+$ \cite{det-CF+_2006} or C$_3$H$^+$ \cite{det-C3H+_2012} 
may provide some clues. The laboratory rotational spectra of C$_2$H$^+$ or C$_5$H$^+$ 
are not known at present, and this applies even to C$_3$H$^+$. The presence of the methoxy 
radical CH$_3$O in the Barnard~1 cloud \cite{det-CH3O_2012} is probably due to PDR activity 
in that area.

Higher angular resolution is able to reduce the line confusion in cases in which more than one 
source are within the beam of a single dish telescope, such as Sgr~B2(N) \cite{det-AAN_2008}. 
Moreover, increased angular resolution will be advantageous for sources which have a small 
angular extension, such as the hot parts ($T > 100$~K) of dense molecular clouds in 
advanced stages of star-formation, so-called hot cores or hot corinos, the CSEs of young or 
of late-type stars, or regions of near-by galaxies or whole galaxies farther away.
A recent molecular line-survey of a $z = 0.89$ foreground galaxy toward the quasar PKS~1830-211 
carried out at 7~mm wavelength with the Australia Telescope Compact Array (ATCA) 
\cite{line-survey_z=0.89_2011} provides a hint of what may be achieved with ALMA or other 
interferometers in terms of Extragalactic sources.

The probably most interesting aspect concerning the astrochemistry of hot cores and 
hot corinos is the issue of complex molecules, which will be discussed separately in 
subsection~\ref{complex-mols}. We suppose that the CSEs of young stellar objects may 
display excitation conditions different from other regions in space, however, the observable 
molecules are likely a subset of those observable in hot cores, hot corinos, or PDRs. 
On the other hand, interferometric observation, in particular those with ALMA, will  
reveal much more of the complex chemistry of CSEs of late-type stars. This subject 
will also be dealt with separately, namely in subsection~\ref{AGB_etc}. Extragalactic 
chemistry usually agrees with certain aspects of Galactic chemistry. However, 
differences in abundance ratios may be significantly large that certain species 
may be observable more easily in Extragalactic sources than in Galactic ones. 
Moreover, sources with significant red-shifts may permit species to be observed 
from ground which cannot be studied in Galactic sources because of transmission 
properties of Earth's atmosphere.

It is worthwhile pointing out that interferometric observations are not always 
advantageous compared to single dish observations as they will resolve out features 
distributed evenly over larger scales. In fact, obtaining a complete picture of the 
distribution of a certain species may require interferometric observations to be 
combined with appropriate single dish data.

%%%%%%%%%%%%%%%%%%%%%%%%%%%%%%%%%%%%%%%%%%%%%%%%%%%%%%%%%%%%%%%%%%%%%%%%%%%%%%%%%%%%%%%%%

\subsection{Considerations for complex molecules}
\label{complex-mols}

The definition of complex molecules in the context of astrochemistry is somewhat loose. 
However, general consensus is that such a molecule should have some complexity both 
structurally as well as spectroscopically, and it should be fairly saturated. 
Hence, fullerenes do not belong to the complex molecules. The minimum number of atoms 
is subject of discussion, usually it is 5 or 6. 

The size of a complex molecule is a very important issue because larger molecules are usually 
less abundant than smaller related molecules. Moreover, larger molecules usually have more 
rotational and vibrational states which may be populated at the elevated temperatures of 
a hot core or hot corino. In addition, larger molecules may have multiple conformers. 
Each aspect contributes to reducing the intensity of an individual line, making ít usually 
much more difficult to detect a larger complex molecule compared with a smaller related 
molecule. Thus, investigations of molecules should focus on those having between 3 and 6 
heavy atoms and taking into account the molecules which have already been detected. 
This does not imply that a moderately large molecule will be found even if a smaller 
related molecule is very abundant in space. On the other hand, a larger molecule not 
related to any molecule found in space thus far may, nevertheless, be identified by 
radio astronomical means.

A considerable number of complex molecules have been detected in space, see e.g. 
\verb"http://www.astro.uni-koeln.de/cdms/molecules". Recent detections include 
aminoacetonitrile \cite{det-AAN_2008}, a potential precursor for glycine in space, 
and ethyl formate and $n$-propyl cyanide \cite{2new-mols_2009}. All three molecules were 
detected in a molecular line survey of Sgr~B2(N) mostly in the 3~mm region and conducted 
with the IRAM 30~m telescope. On the other hand, glycine, the simplest amino acid, 
has not yet been detected in space \cite{not-yet-glycine_2005}. 

There have been very extensive and diverse laboratory investigations into the spectroscopy 
of complex molecules such that it is not possible to provide a comprehensive overview here. 
However, we will give some general considerations and some examples which will permit to 
evaluate data needs of ALMA and other interferometers. 

The recent detection of $n$-propyl cyanide ($n$-C$_3$H$_7$CN) in space \cite{2new-mols_2009} 
prompted investigations into the rotational spectra of $i$-propyl cyanide \cite{i-PrCN_rot_2011} 
and of $n$-butyl cyanide \cite{n-BuCN_rot_2012}, permitting to study the ratio of branched 
versus straight-chained molecules and to search for the next larger homolog, respectively. 
And with data on $c$-propyl cyanide available \cite{c-PrCN_rot_2008}, more insight into 
the importance of cyclic species is possible. The rotational spectrum of ethyl cyanide 
has been studied recently \cite{EtCN_rot_2009} as well as several isotopologs of 
methyl cyanide \cite{MeCN_rot_2009} and several vibrational states of vinyl cyanide 
\cite{VyCN_rot_2012}. Moreover, sufficient data should be available for both C$_3$H$_7$CN 
isomers; but some of the C$_3$H$_5$CN isomers may require additional spectroscopic 
investigations.

Soon after the detection of aminoacetonitrile, the rotational spectrum of 2-aminopropionitrile
has been studied rather extensively \cite{2-APN_rot_2012}, however, only sparse data are 
available for 3-aminopropionitrile. And even though the data set of methyl acetate, an isomer 
of ethyl formate, has been extended considerably recently \cite{MeAc_rot_2011}, it is probably 
still not sufficient for a detection in space. However, observations of transitions extending 
to higher quantum numbers are under way.

It is important to point out that for the more abundant molecules in space it is not sufficient 
to investigate the vibrational ground state of the main isotopic species. Minor isotopic 
species, in particular those involving $^{13}$C or D, may be observable or transitions 
belonging to excited vibrational states. In fact, even though several excited vibrational 
states of ethyl cyanide are being studied, and sufficient data have been published for the 
ground vibrational states of all singly substituted isotopologs, at least the lowest two, 
maybe five excited vibrational states of the singly substituted $^{13}$C isotopic species will 
be observable with ALMA and probably also the ground states of the doubly substituted species 
because the $^{12}$C/$^{13}$C ratio in Galactic center sources is as low as 20. 

Methanol is probably the most important complex molecule in space judged by its great abundance 
in various astronomical sources and its many transitions covering frequencies from the microwave 
region well into the terahertz region. The ground and the first two torsional states have been 
studied rather extensively \cite{MeOH_rot_2008}, but this may still be insufficient for ALMA, 
even if only these three vibrational states are considered. Additional data are definitively 
needed for higher vibrational states. Very recently, extensive data have been published on 
CH$_2$DOH, \cite{CH2DOH_rot_2012}, and several other isotopic species are under investigation. 
CHD$_2$OH is one isotopolog for which the data situation is insufficient and which is  
currently, to the best of our knowledge, not under investigation.

The ground vibrational state of ethanol has been studied very extensively \cite{EtOH_rot_2008}, 
but for low-lying excited vibrational states, which may be observable with ALMA, only 
very limited data have been published. Information is available on the lowest energy conformer 
of each of the two $^{13}$C isotopomers \cite{13C-EtOH_rot_2012}. Future ALMA observations 
will show if the amount of data is sufficient. The singly deuterated species are currently 
under investigation.

There is usually a sharp drop in column densities from methanol to ethanol. Therefore, it is 
not surprising that $n$-propanol has not yet been detected despite extensive laboratory 
work \cite{n-PrOH_rot_2010}. Nevertheless, it is quite likely that the molecule will be 
detected with ALMA or maybe other instruments eventually, but the column densities of 
$n$-butanol may be so low that it may never be detectable in space.

Only very few reports exits which deal with the presence of methyl mercaptan, CH$_3$SH, 
the sulfur analog of methanol, in space. The recent study on its rotational spectrum up 
to the second torsional state \cite{CH3SH_rot_2012} may help to change this situation. 
Ethyl mercaptan is currently under investigation. The prospects of detecting this molecule 
in space are uncertain.

The need to consider many vibrational states for some of the complex molecules in combination 
with frequent vibration-rotation interaction, which can make the analyses of rotational 
spectra very demanding, have lead to an alternative proposal to analyze the occurance of 
these molecules in space \cite{labspec-T_2007}. Rotational spectra are recorded quantitatively 
at very many different temperatures and over large frequency ranges. This method has been 
tested very recently for the analysis of ALMA science verification observations of Orion~KL 
\cite{ALMA_lab-spec-T_2012}. One needs to point out that extrapolation in frequency is 
impossible, and extrapolation in temperature is difficult, in particular to higher temperatures. 
In addition, the method requires local thermodynamic equilibrium (LTE) to be a good assumption 
$-$ at least for the respective molecule in that very source. Future studies will show how 
useful this method is especially for the study of complex molecules in hot core or 
hot corino sources.

It should be pointed out that some O-rich complex molecules have been found in Galactic center 
sources at rather low rotational temperatures \cite{cold_complex_mols_2006,cold_complex_mols_2008}. 
In the course of the review process of this article, the detections of cyanomethanimine 
\cite{det-cyanomethanimine_2013} and ethanimine \cite{det-ethanimine_2013} toward Sgr~B2(N) 
have been reported employing the Green Bank 100~m telescope (GBT). The very low rotational 
temperatures suggest that these molecules are released into the gas phase by physical processes 
similar to those of the O-rich complex molecules. It may well be that such molecules are 
not easily detectable for interferometers because of their spatially extended occurance. 
They are more likely detected with single dish telescopes; large dishes, such as the GBT, 
are necessary for the detection of heavier molecules at longer wavelengths. 

%%%%%%%%%%%%%%%%%%%%%%%%%%%%%%%%%%%%%%%%%%%%%%%%%%%%%%%%%%%%%%%%%%%%%%%%%%%%%%%%%%%%%%%%%

\subsection{Considerations for circumsteller envelopes of late-type stars}
\label{AGB_etc}

After a star has consumed the hydrogen in its innermost region, it starts to convert helium 
into carbon, which in turn will be converted to heavier elements such as oxygen and neon. 
The outer part of the star expands greatly, and the star is transformed into an Asymptotic 
Giant Branch (AGB) star. These as well as other late type stars eject large amounts of 
gas and dust, creating a circumstellar envelope. The chemistry of late type stars depends 
very much on the ratio of oxygen to carbon, because these are the two most abundant elements 
in space after hydrogen and helium, and by far the most of carbon \textit{or} oxygen is 
locked up in CO. The CSEs of O-rich (M-type) stars turn their oxygen excess into H$_2$O, SiO, 
SO, SO$_2$, and other, mostly simple oxides. The CSEs of C-rich (C-type) stars, in contrast, 
show a large variety of molecules consisting to a large extent of one or more carbon atoms 
(carbon-chain molecules). Late-type stars with approximately equal amounts of oxygen and 
carbon in their outer parts are classified as S-type stars.

CW~Leonis, or short just CW~Leo, also known as IRC~+10216, is the most prominent C-rich 
late-type star. Until quite recently, it was thought that CW~Leo would be a very unusual 
AGB star because of the very rich chemistry in its CSE. It turned out, however, that it is 
a fairly typical C-type AGB star. The ease of detecting so many different molecules is simply 
caused by its proximity. Among recent noteworthy detections are those of anions C$_6$H$^-$, 
C$_4$H$^-$, C$_8$H$^-$, C$_3$N$^-$, C$_5$N$^-$, and most recently CN$^-$ \cite{det-CN-_2010}. 
Other detections include CCP \cite{det-CCP_2008}, KCN \cite{det-KCN_2010}, and FeCN 
\cite{det-FeCN_2011}. The unusually large number of SiC$_2$ transitions detected with 
Herschel \cite{SiC2_2012} has already been mentioned earlier.

In recent years, the chemistry of oxygen-rich late-type stars turned out to be richer than 
initially thought. VY~Canis Majoris, or short just VY~CMa, is a variable star of the Mira 
type, and because of its very large mass loss the most prominent O-rich late-type star. 
In fact, PO \cite{det-PO_2007}, AlO \cite{det-AlO_2009}, and AlOH \cite{det-AlOH_2010} 
were detected for the first time and thus far only in the CSE of VY~CMa.

Based on the already detected molecules, one would expect further metal oxides and hydroxides 
to be detected in the envelopes of O-rich stars, sulfides and hydrosulfides should be much 
less abundant, but may still be found. Further anions, e.g. CH$_2$CN$^-$ or related species, 
further metal cyanides or isocyanides, and molecules containing P or Si may be observable 
in the envelopes of C-rich stars. Metal fluorides and chlorides may be found in the CSEs of 
all types of late-type stars. Some transition metal hydrides have been detected in the 
atmospheres of rather cool stars employing electronic spectroscopy. Very high angular 
resolution is probably required to detect such molecules by means of radio astronomy.

There is a demand for new or additional measurements for several molecular species even 
if rotational spectra have been studied for quite a few of these. Recent investigations 
include those on H$_2$SiS \cite{H2SiS_rot_2011}, OSiS \cite{OSiS_rot_2011}, 
CCP \cite{CCP_rot_2009}, PCN \cite{PCN_rot_2012}, TiS \cite{TiS_rot_2010}, 
CrS \cite{CrS_rot_2010}, FeCN and FeNC \cite{FeCN_FeNC_rot_2011}, and 
MnH \cite{MnH_rot_2008}. In light of the recent investigations of the rotational 
spectrum of TiO$_2$ \cite{TiO2_rot_2008,TiO2_rot_2011}, it will be interesting to know 
if metal dioxides or dihalides may be detectable in space. In fact, in the course of the 
review process of this article, the detection of TiO$_2$ in the circumstellar envelope 
of VY~CMa has been reported using the Submillimeter Array (SMA) and the Plateau de~Bure 
Interferometer (PdBI) \cite{TiO_TiO2_VY-CMa_2013}. This observation, as well as similar 
ones in the future, will shed light on the dust formation process which is far from 
understood at present.

%%%%%%%%%%%%%%%%%%%%%%%%%%%%%%%%%%%%%%%%%%%%%%%%%%%%%%%%%%%%%%%%%%%%%%%%%%%%%%%%%%%%%%%%%
%%%%%%%%%%%%%%%%%%%%%%%%%%%%%%%%%%%%%%%%%%%%%%%%%%%%%%%%%%%%%%%%%%%%%%%%%%%%%%%%%%%%%%%%%
%%%%%%%%%%%%%%%%%%%%%%%%%%%%%%%%%%%%%%%%%%%%%%%%%%%%%%%%%%%%%%%%%%%%%%%%%%%%%%%%%%%%%%%%%

% \section*{Acknowledgements}
\ack
We are grateful to Peter Schilke for providing Figure~1. 
HSPM acknowledges support by the Bundesministerium f\"ur Bildung und Forschung (BMBF) 
through project FKZ~50OF0901 (ICC HIFI {\it Herschel}).
CPE was supported within the VAMDC project which was funded under the ''Combination 
of Collaborative Projects and Coordination and Support Actions'' Funding Scheme of 
The Seventh Framework Programme. Call topic: INFRA-2008-1.2.2 Scientific Data 
Infrastructure. Grant Agreement number: 239108. Recent funding has been provided 
by the BMBF through the German ALMA Regional Centre (ARC) Node via project number 
5A11PK3.

\end{document}